\documentclass[letter]{aa} 

\usepackage{graphicx}
\usepackage{hyperref}
\usepackage{txfonts}
\newcommand{\Msolar}{M_{\odot}}
\newcommand{\frb}{{FRB\,20201124A}}
\newcommand{\frba}{{FRB\,20121102A}}
\newcommand{\frbb}{{FRB\,20180916B}}

\begin{document} 

\title{
The Fast Radio Burst FRB 20201124A in a star forming region: constraints to the progenitor and multiwavelength counterparts }

   \subtitle{}

   \author{L. Piro
          \inst{1}
          \and
          G. Bruni
          \inst{1}
          \and
          E. Troja \inst{2,3}
          \and
          B. O'Connor \inst{2,3,4,5}
          \and
          F. Panessa \inst{1}
          \and
          R. Ricci \inst{6,7}
          \and
          B. Zhang \inst{8}
          \and
          M. Burgay \inst{9}
          \and
          S. Dichiara \inst{2,3}
          \and
          K. J. Lee \inst{10}
          \and
          S. Lotti \inst{1}
          \and
          J. R. Niu \inst{11}
          \and
          M. Pilia \inst{9}
          \and 
          A. Possenti \inst{9,12}
          \and
          M. Trudu \inst{9,12}
          \and
          H. Xu \inst{10}
          \and
          W. W. Zhu \inst{11}
          \and
          A. S. Kutyrev \inst{2,3}
          \and
          S. Veilleux \inst{2}
          }
          



   \institute{INAF -- Istituto di Astrofisica e Planetologia Spaziali, via Fosso del Cavaliere 100, 00133 Rome, Italy\\
              \email{luigi.piro@inaf.it}
         \and
Department of Astronomy, University of Maryland, College Park, MD 20742-4111, USA 
\and
Astrophysics Science Division, NASA Goddard Space Flight Center, 8800 Greenbelt Rd, Greenbelt, MD 20771, USA 
  \and
  Department of Physics, The George Washington University, 725 21st Street NW, Washington, DC 20052, USA
  \and
Astronomy, Physics, and Statistics Institute of Sciences (APSIS), The George Washington University, Washington, DC 20052, USA
\and
 INAF -- Istituto di Radioastronomia, Via Gobetti 101, I-40129, Bologna, Italy 
 \and
Istituto Nazionale di Ricerca Metrologica (INRiM) - Strada delle Cacce 91, I-10135 Torino, Italy
\and
Department of Physics and Astronomy, University of Nevada, 89154, Las Vegas, NV, USA
\and
INAF -- Osservatorio Astronomico di Cagliari, via della Scienza 5, I-09047 Selargius (CA), Italy
\and 
Kavli Institute for Astronomy and Astrophysics, Peking University, Beijing 100871, China
\and
National Astronomical Observatories, Chinese Academy of Sciences, Beijing 100012, China
\and
Department of Physics, Universit\`a di Cagliari, S.P. Monserrato-Sestu km 0,700, I-09042 Monserrato (CA), Italy
             }

   \date{Received; accepted:}

\abstract
{We present the results of a multiwavelength campaign of \frb, the  third closest repeating fast radio burst recently localized in a nearby ($z$\,$=$\,$0.0978$) galaxy. 
 Deep VLA observations led to the detection of quiescent radio emission, also marginally 
 visible in X-rays with \textit{Chandra}. Imaging at 22 GHz allowed us to  resolve the source on a scale of $\gtrsim 1\arcsec$ and locate it at the position of the FRB, within an error of $0.2\arcsec$. EVN and e-MERLIN observations sampled small angular scales, from 2 to 100 mas, providing tight upper limits on  the presence of a compact source and evidence for diffuse radio emission. We argue that this emission is associated with enhanced star formation activity in the proximity of the FRB, corresponding to a star formation rate of $\approx 10\ {\rm M}_\odot {\rm yr}^{-1}$. The surface star formation rate at the location of \frb\ is two orders of magnitude larger than typically observed in other precisely localized FRBs.
Such a high SFR is indicative of this FRB source being a new-born magnetar produced from a SN explosion of a massive star progenitor. Upper limits to the X-ray counterparts of 49 radio bursts observed in our simultaneous FAST, SRT and \textit{Chandra} campaign are consistent with a magnetar scenario.
}
               \titlerunning{FRB\,20201124A}
               \authorrunning{L. Piro et al.}
   \maketitle
%

\section{Introduction}
Fast radio bursts (FRB) are bright radio flashes of ms duration \citep{Lorimer07,Petroff19,Cordes19}, located at cosmological distances. 
Hundreds of these events have now been discovered \citep{chime21}, yet their progenitor systems remain unknown. 
A leading candidate source is a highly magnetized neutron star (NS), known as magnetar \citep{Popov10,Katz16,Metzger17,beloborodov17,Kumar17, Wadiasingh19, Lyubarsk21,Yang21}, as supported by the recent detection of FRB-like radio bursts from a soft gamma-ray repeater in our galaxy \citep{chimesgr,Bochenek20,Li21,Mereghetti20}. 
However, the heterogeneous environment of FRBs and their empirical distinction in two classes (repeaters and non-repeaters) leave the door open to multiple progenitor channels \citep{Zhang20}.

  
 Accurate and rapid localizations of FRBs are key to trigger  multiwavelength follow-up observations, which provide vital information on the FRB host galaxy, its distance scale, 
 the broad band spectrum of the burst and the presence on any quiescent emission associated to the flaring source.
  However, FRBs span a wide range of distances, and the majority are too far to place any meaningful constraints on persistent emission or associated X-ray bursts. Therefore, we organized a follow-up program to focus on the closest events ($z$\,$\lesssim 0.1$) 
  localized to an accuracy of at least $\lesssim10\arcsec$.

Starting from March 21, 2021 the repeating source \frb\ was reported to be in an active state \citep{ChimeAtel21,askapdet}, with continued activity over the next few months.
We initiated our multiwavelength campaign soon after the FRB localization and the  unambiguous identification of its host galaxy \citep{Day21,Law21,Wharton21}
that, at a redshift $z$\,$\sim$\,$0.098$ \citep{Kilpatrick21}, meets our selection criteria.

In this paper we present (\S\ref{sec: obsmain}) the results of our multi-instrument observations of \frb\ in radio with the Very Large Array (VLA), the European VLBI Network (EVN), the enhanced Multi Element Remotely Linked Interferometer Network (e-MERLIN), the Five-hundred-meter Aperture Spherical radio Array (FAST), the Sardinia Radio Telescope (SRT), the upgraded Giant Metrewave Radio Telescope (uGMRT), in X-ray with the \textit{Neil Gehrels' Swift Observatory} (\textit{Swift})  and the {\it Chandra X-ray Observatory} (CXO), and in the optical with the Lowell Discovery Telescope (LDT).
In \S\ref{sec: sfrmain} we present the detection of a quiescent source associated to \frb\ in radio and X-rays and discuss its origin. In \S\ref{sec: interpmain} we discuss our results in the context of the properties of the local and host galaxy environments and their implications for the progenitors of FRBs, with particular regard to the magnetar scenario.
In \S\ref{sec: limitsmain} we present the results of the simultaneous observation of FAST, SRT and \textit{Chandra}, aimed at detecting X-ray counterparts of radio bursts, and discuss implications on the central engine. We summarize our conclusions in \S\ref{sec: concmain}.


\begin{figure*}
\centering
 \includegraphics[width=0.97\columnwidth]{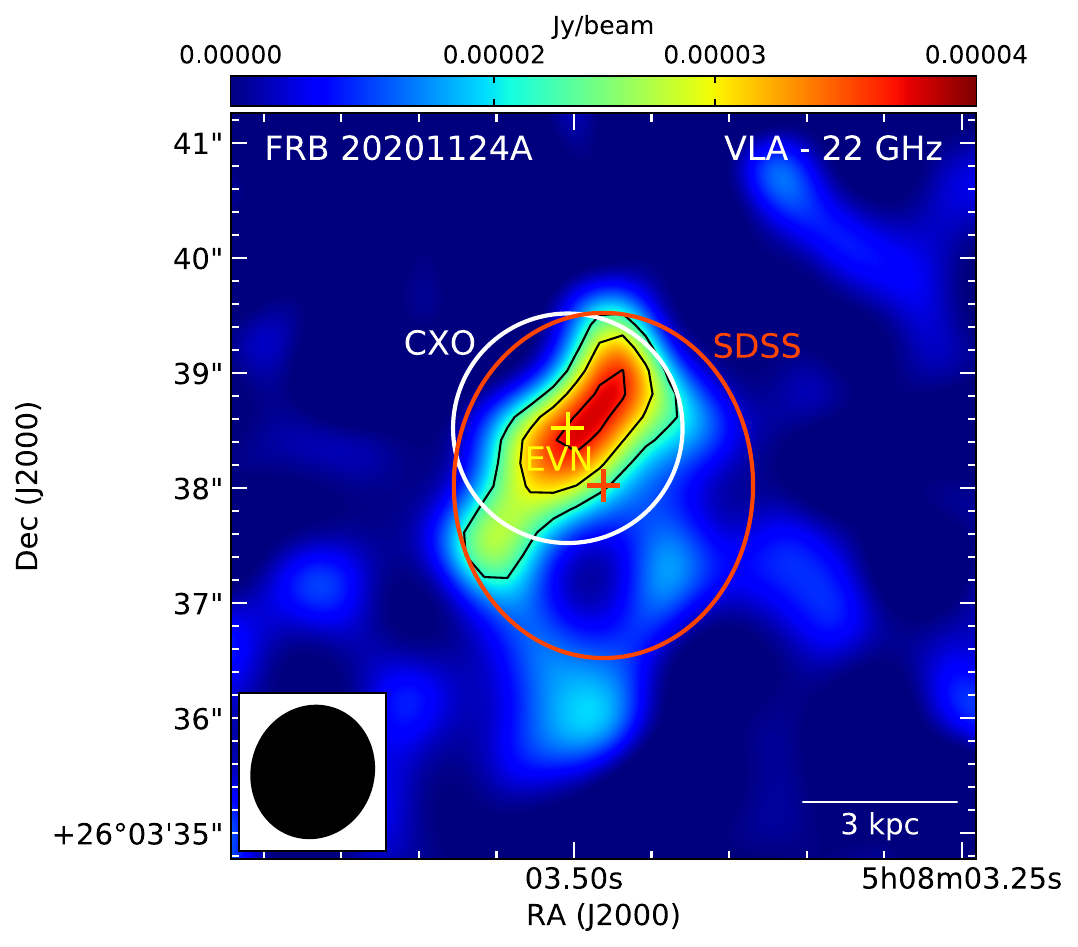}
  \includegraphics[width=0.97\columnwidth]{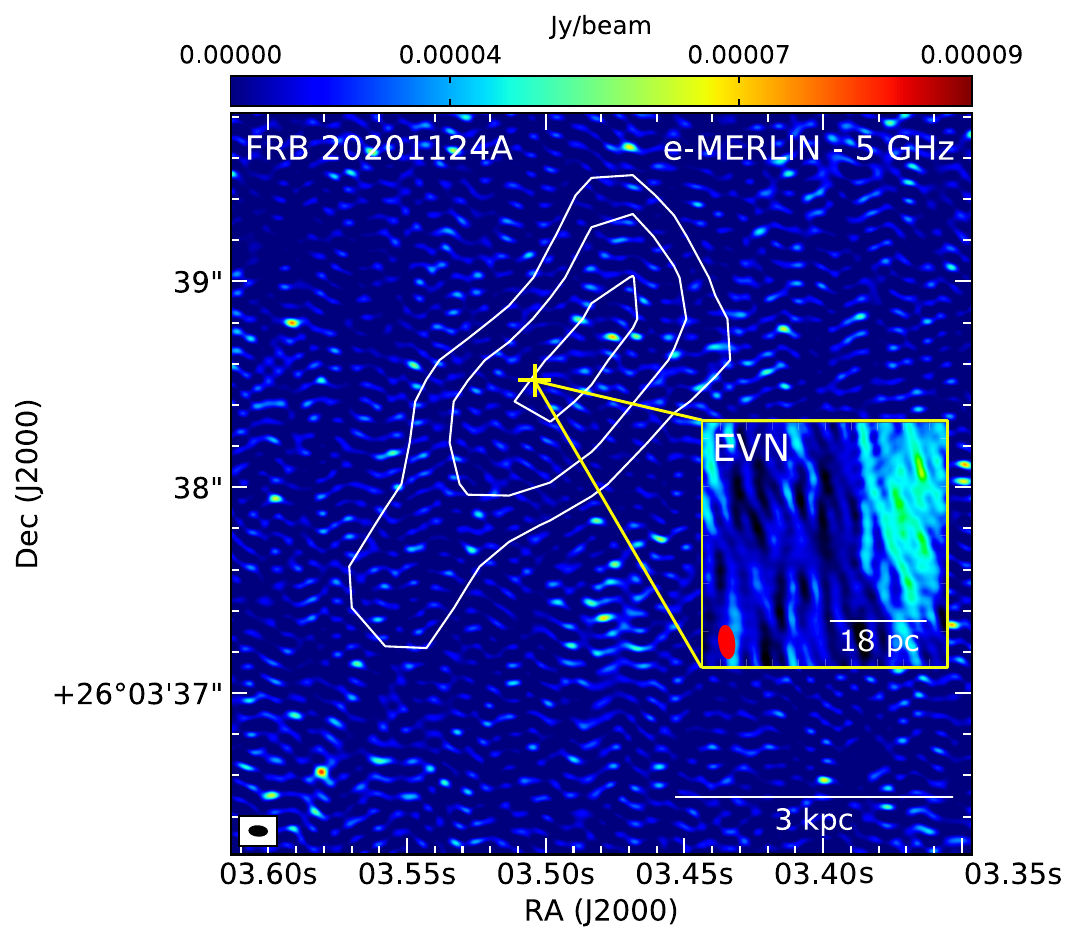}
\caption{Imaging and location of \frb\ from radio to X-rays.  \textbf{Left panel:} The VLA image at 22 GHz presented in this work (in colors), with contours indicating the 3, 4, and 5 sigma significance levels. The FWHM is reported in the lower-left corner; SDSS host galaxy centroid and half-light radius (red); CXO extraction region (white), centered on the EVN FRB position (yellow cross)
\textbf{Right panel:}
e-MERLIN image at 5 GHz of the persistent radio source region, showing no detection. Contours are from the VLA image at 22 GHz. The cross indicates the location of \frb\ from EVN observations \citep{Marcote21}. The FWHM is reported in the lower-left corner. The inset shows the EVN image from our campaign (25$\times$25 mas) with no detection at the FRB location, also at 5 GHz.
}
\label{fig:image}
\end{figure*}

\section{Observations}
\label{sec: obsmain}

\subsection{Radio}
We searched for  a persistent radio source associated to the FRB with the 
Jansky Very Large Array (JVLA). Observations
were carried out under program (SG9112; PI: Piro) at multiple frequencies between
April 9 and June 22 (see Appendix \ref{sec:VLA}). 
A radio source, consistent with the FRB position, was detected  with a flux of $(340\pm30)~\mu{\rm Jy}$ at 3 GHz (\cite{Ricci21}, see also \cite{Wharton_21b,Ravi21}).   VLA monitoring of this source
revealed that the radio emission is not variable, but persistent. 
However, the initial VLA D configuration did not have the sufficient resolution to resolve its angular extent. 
Therefore, we imaged the field using various beam sizes with  the EVN ($\approx$ 2 mas), e-MERLIN ($\approx 50-100$ mas) and VLA ($\approx$\,1\arcsec). 
Our EVN and e-MERLIN campaigns (Appendix \ref{sec:EVN}) did not detect any compact source over angular scales ranging from 2 mas to 60 mas. At the host redshift of $z$\,$\approx$0.098, this corresponds to a projected linear size from 3 to 110 pc, and implies that the persistent radio emission is extended with an angular size $\gtrsim 0.1$\arcsec, as also independently reported  by \citet{Marcote21,Ravi21}. 

High-frequency (22~GHz) VLA observations, carried out in C configuration,  partially resolve the source at the scale of $\gtrsim$1\arcsec. As shown in Fig.~\ref{fig:image}, 
the centroid of the radio emission
coincides with the precise FRB position derived by EVN observations \citep{Marcote21} within our location accuracy of $0.2\arcsec$.
The radio source is elongated  over $\approx 2\arcsec$
 at a position angle of 140$^\circ$. Its centroid displays marginal evidence of an offset of $0.42\arcsec$ from the galaxy center, more evident when  the peak of the radio emission (5 sigma level) is considered.
At the redshift of  $z$=0.0978 (see Appendix \ref{sec: opt_spectroscopy}), the offset of the radio source from the galaxy center  would correspond to   0.8 kpc and its extension to $\approx$\,3 kpc.

To extend the frequency coverage at low frequencies, we carried out uGMRT observations on July 21 and 24, 2021 (project code ddtC194; PI: Bruni), using the band-3 (260-500 MHz) and band-2 (125-250 MHz) receivers, respectively. We detected the source as an unresolved component with a flux density of 1.7$\pm$0.2 mJy at 380 MHz. The summary of our radio observations is in Tab.\ref{tab:radio}.

\subsection{Optical}

Observations with the Ultra-Violet Optical Telescope (UVOT) on-board \textit{Swift} were performed on April 6, 2021 (PI: Piro, Target ID: 14258) in $u$-band with a total exposure 9.9 ks. We utilized the \texttt{uvotimsum} task within \texttt{HEASoft v.~6.27.2} to coadd multiple exposures, and extracted the photometry with the \texttt{uvotsource} tool using a circular aperture of $3\arcsec$ radius. In the stacked image, we derive an upper limit $u\gtrsim 22.8$ AB mag.

On April 21, 2021 we obtained optical spectroscopy of the putative host galaxy using the DeVeny spectrograph mounted on the 4.3m Lowell Discovery Telescope (LDT) for a total exposure of 4$\times$600 s. DeVeny was configured with the 300 g mm$^{-1}$ grating and a 1.5\arcsec slit width,
covering the FRB location.
The spectrum covers wavelengths $3600\, \AA$\,-\,$8000\, \AA$ at a dispersion of $2.2\,\AA$ pix$^{-1}$.
The resulting spectrum, derived as described in Appendix \ref{sec: opt_spectroscopy}, is displayed in Figure \ref{fig:host_spectrum}.
Emission lines (Tab.\ref{tab:emlines}), indicative of on-going star-formation, are detected at $\lambda_\textrm{obs}\approx 7207$, $7230$, $7374$, and $7391\,\AA$ which we associate to H$\alpha$, the [NII] doublet, and the [SII] doublet at a redshift $z=0.0978 \pm 0.0002$ consistent with previous estimates \citep{Ravi21,Fong21}. 


\subsection{X-ray}
\label{sec: xraymain}
Observations with the \textit{Swift} X-ray Telescope (XRT) were carried out as part
of the \textit{Swift} Guest Investigator Program (PI: Piro). 
Data were collected in Photon Counting (PC) mode starting  
on April 6, 2021
for a total of 9.9 ks exposure (Appendix \ref{sec:Chandraimage}). 
No source was detected at the location of the FRB
down to a 3~$\sigma$ upper limit $<1.3\times 10^{-3}$ cts s$^{-1}$ (0.3-10~keV). We convert this value into an unabsorbed flux $<1.0\times 10^{-13}$ erg cm$^{-2}$ s$^{-1}$ (0.3-10~keV) assuming an absorbed power-law with photon index $\Gamma=2.0$ and a Galactic hydrogen column density $N_H=4.5\times 10^{21}$ cm$^{-2}$ \citep{Willingale2013}. 

Deeper imaging of the field was carried out with the
\textit{Chandra X-ray Observatory} (CXO) observations under Director's Discretionary Time (DDT). The observations (ObsID: 25016; PI: Piro) occurred on April 20, 2021 for a total of 29.7 ks. 
Our analysis (Appendix \ref{sec:Chandraimage}) was performed in the $0.5-7$ keV energy range.
A blind search with \texttt{wavdetect} did not find any source at the FRB position,
however a targeted search revealed a weak X-ray detection.  
We used a circular source region of 1\arcsec radius centered at the EVN position \citep{Marcote21} and, within this region, measured a total of 3 photons. 
We estimate a background level of $\lesssim$0.25 cts from nearby source-free regions and a source significance of 99.7\%  \citep[3~$\sigma$ Gaussian equivalent;][]{Kraft1991}.

The observed 0.5-7.0 keV count-rate 1.0$^{+0.7}_{-0.6}$ $\times$ 10$^{-4}$ cts s$^{-1}$ is 
converted into flux using the same absorbed power-law model described above. 
For a photon index $\Gamma=2.0$, the unabsorbed flux in the 2-10~keV band is 
$\rm{F_X}=  1.3^{+0.9}_{-0.8} \times 10^{-15}$ erg cm$^{-2}$ s$^{-1}$ corresponding to a luminosity of 
$\rm{L_X=(3.1 \pm 1.9 )\ 10^{40} erg \ s^{-1}}$.


\subsection{FRB searches}

\begin{figure}
\includegraphics[trim= 110 50 110 100, clip, width=
1.4\columnwidth,origin=c]{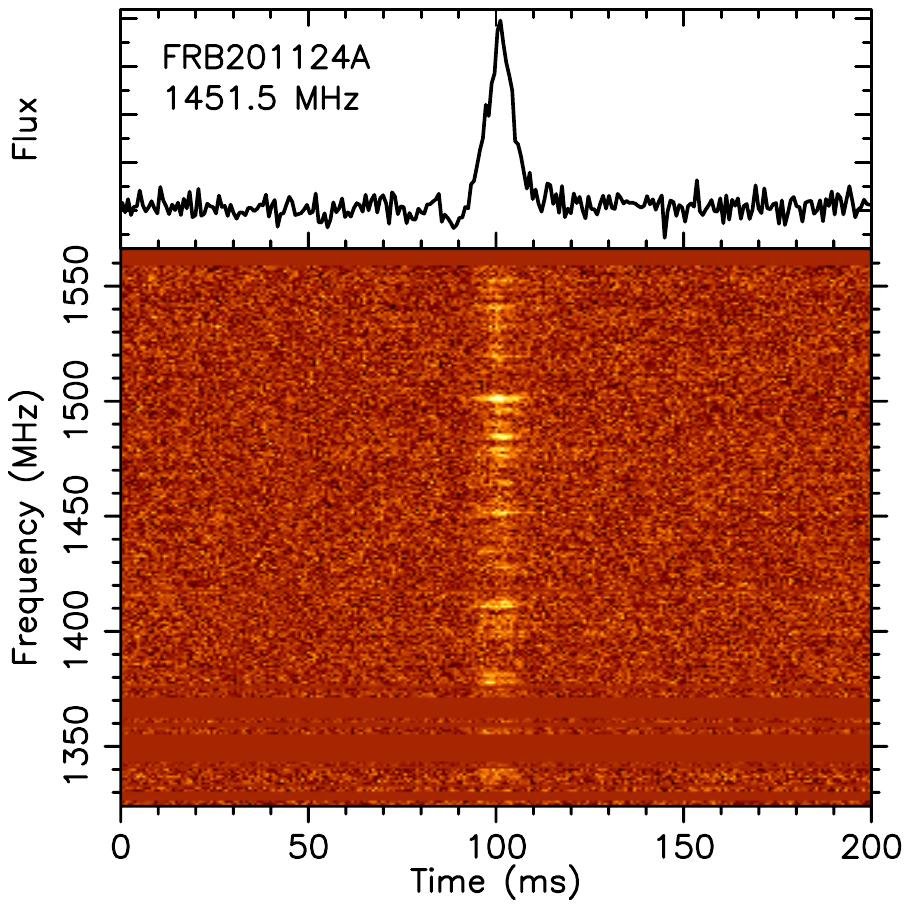}
\\
\includegraphics[width= \columnwidth]{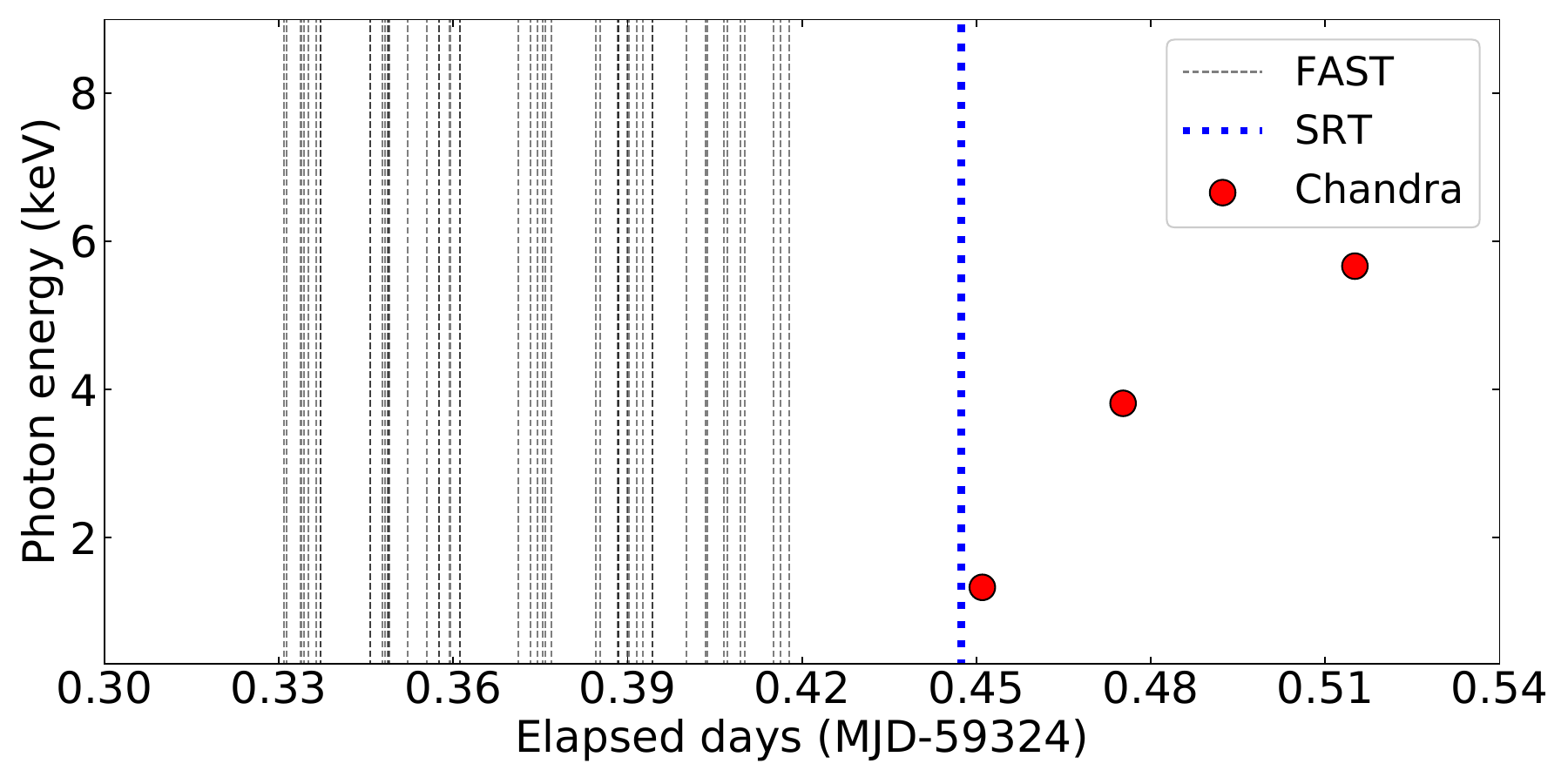}
\caption{Search for X-ray counterparts of bursts from \frb. \textbf{Top panel:} The brightest FRB  simultaneous with our \textit{Chandra} observation was observed by SRT at T  = 59324.44738212(12) MJD  (reported at infinite frequency in Barycentric Dynamical Time units), with a width= 10 ms, a fluence = 13 Jy ms, and
DM = $421\pm4$ pc cm$^{-3}$.  
\textbf{Bottom panel}: Arrival times of FRBs detected by FAST (gray lines) and SRT (blue line) compared to the arrival times of \textit{Chandra} photons (red circles) coincident with the FRB position. The X-ray event closest to an FRB occurs 314 s after the SRT detected FRB. 
}
 \label{fig:FRB_SRT}
\end{figure}

We have  organized a simultaneous coverage of the \textit{Chandra} observation  with FAST and SRT  aimed at searching for X-ray counterparts of   bursts from \frb (Appendix \ref{sec:FRB}).
The  radio facilities covered two separated   periods for a total of 60\% of the 8.3 hour {\it Chandra} observation. A total of 49 radio bursts (48 by FAST, 1 by SRT), with fluences in the range 0.017- 13 Jy ms and with an average duration of about 19 ms were observed.  We do not find any significant coincidences  with the 3 X-ray photons detected by \textit{Chandra}. The closest FRB-X-ray photon time difference is 314 s. (Fig.\ref{fig:FRB_SRT}).


\begin{figure}
\centering
\includegraphics[width=\columnwidth]{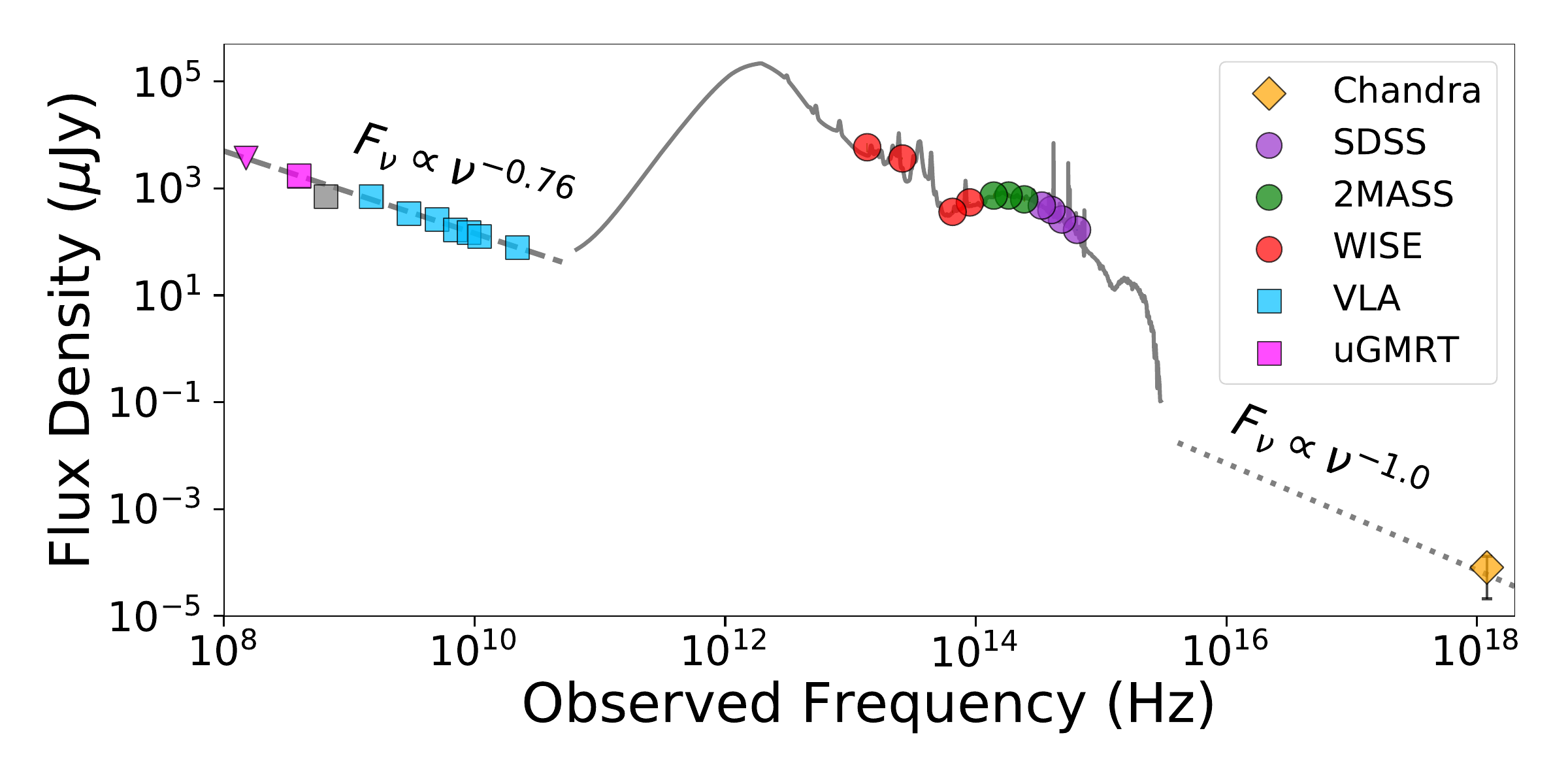}

    \caption{Broadband radio to X-ray spectral energy distribution of \frb. We show the best fit models to the radio (Appendix \ref{sec:radiospectrum}), optical and infrared (Appendix \ref{sec: Galaxy SED modeling}), and X-ray data (Appendix \ref{sec:Chandraimage}) in gray. The dotted line represents the X-ray spectrum expected from the star formation derived from the optical SED modeling. It is dominated by a  population of HMXB with a spectrum $\approx \nu^{-1}$ \citep{Fragos2013}. 
    The downward magenta triangle represents the uGMRT upper limit at 200 MHz. The gray square represents the uGMRT data at 650 MHz published by \citet{Wharton_21b}.
    }
    \label{fig:broadband_SED_wRadio}
\end{figure}


\section{Origin of the radio emission}
\label{sec: sfrmain}
We argue that the extended  radio source is associated  to a star forming region in the host galaxy and  exclude an AGN origin, as discussed in Appendix \ref{sec:persistentorigin} \citep[see also][] {Ravi21, Fong21}.
Radio emission from star formation in normal and starburst galaxies is due to the combination of synchrotron emission from cosmic ray electrons produced in supernova remnants, that dominates below $\approx 20$ GHz, described by a power law with a slope $\alpha_S$, and a flat free-free emission and its associated low frequency absorption $\nu_{ff, abs}$ (see Appendix \ref{sec:radiospectrum}). 
In our source, the spectrum from 0.15 GHz to 22 GHz is well described by a power law with $\alpha_S=0.76\pm0.07$, with no evidence of low frequency absorption or thermal emission (Fig.\ref{fig:broadband_SED_wRadio}). Such a spectrum is consistent with that observed in star-forming galaxies \citep{Tabatabaei17, Klein18}. 
The star formation rate associated to this source can be derived using the standard  correlation relating SFR to radio luminosity, SFR= $\frac{L_{1.4GHz}}{1.6\times 10^{28}{\rm erg s^{-1} Hz^{-1}}}{\rm M_\odot yr^{-1}}$ \citep{Murphy11}. From the observed flux (Tab.\ref{tab:radio}) and the corresponding luminosity  L(1.4GHz)=$(1.7\pm0.2)\times 10^{29} {\rm erg\  s^{-1} Hz^{-1}}$ we derive  SFR$\approx 10 {\rm M_\odot yr^{-1}}$.

The X-ray data are also in agreement with emission from a star forming region.
X-ray emission is  expected from the population of low mass (LMXB) and high-mass X-ray binaries (HMXB). By adopting standard correlations linking SFR to X-ray luminosity \citep{Fragos2013,Ranalli2003} one expects a 2-10 keV luminosity  in the range ${\rm L_X}\approx(3-10) \times 10^{40}\ {\rm erg\ s^{-1}}$, consistent with the \textit{Chandra} measurement (\S\ref{sec: xraymain} and Appendix \ref{sec:Chandraimage}).

Consistent values of the SFR and the key properties of the host galaxy are derived from optical spectroscopy and the spectral energy distribution (SED) modeling (Appendices \ref{sec: opt_spectroscopy} and \ref{sec: Galaxy SED modeling}).
The best fit parameters describing the galaxy are: a high intrinsic extinction $A_V=1.3 \pm 0.2$ mag, symbolic of a dusty galaxy, a sub-solar metallicity $Z_\odot=0.46 \pm 0.25$, an $e$-folding time $\tau=4.3^{+3.6}_{-2.5}$ Gyr, a moderate stellar mass $\log(M_*/M_\odot)=9.78 \pm 0.08$, an old stellar population $t_m=1.5^{+0.5}_{-0.4}$ Gyr, and a star-formation rate SFR $=4.3 \pm 0.5$ $M_\odot$ yr$^{-1}$.
The best fit model spectrum is shown in Fig.\ref{fig:broadband_SED_wRadio}.
From optical spectroscopy we derive a redshift $z=0.0978 \pm 0.0002$ and  an ${\rm H\alpha}$ line luminosity,  ${\rm L(H\alpha)= (4.5\pm0.8)\times 10^{41} erg\ s^{-1}}$, corresponding to   SFR$_{\textrm{H}\alpha}=2.3\pm0.4$ $M_\odot$ yr$^{-1}$. As discussed in Appendix \ref{sec: opt_spectroscopy}, the $\textrm{H}\alpha$ luminosity can underestimate the total SFR for observational reasons. Physical reasons can further reduce the SFR$_{\textrm{H}\alpha}$ compared to the radio-derived SFR. Radio  is also sensitive  to  heavily obscured star-formation and, in addition, it captures longer star-forming ages, up to 100 Myr \citep{Condon92}, whereas optical traces recent star formation, up to 10 Myr. 
The mismatch between optical (low SF) and radio (high SF)  may  be interpreted as  a contribution from heavily obscured SF or as a post starburst episode, that is 100 Myr ago the SFR was as high as 10 M$_{\odot}$yr$^{-1}$ but it rapidly declined and within 10 Myr was less than 2 M$_{\odot}$yr$^{-1}$.  A  decrease of the SFR at a recent age is also suggested  by a non parametric modelization of the star forming history \citep{Fong21}, indicating that the SFR was higher in the past galaxy's history (although at a value not as large as the radio-derived SFR) before decreasing by a factor of two  $\approx$ 30 Myr ago.
We conclude that the properties of the radio, optical and  X-ray emission are consistent with those expected by star formation. 

   \section{The local star forming  environment as a clue to the progenitor}
   \label{sec: interpmain}
    
\begin{figure}
 \includegraphics[width=\columnwidth]{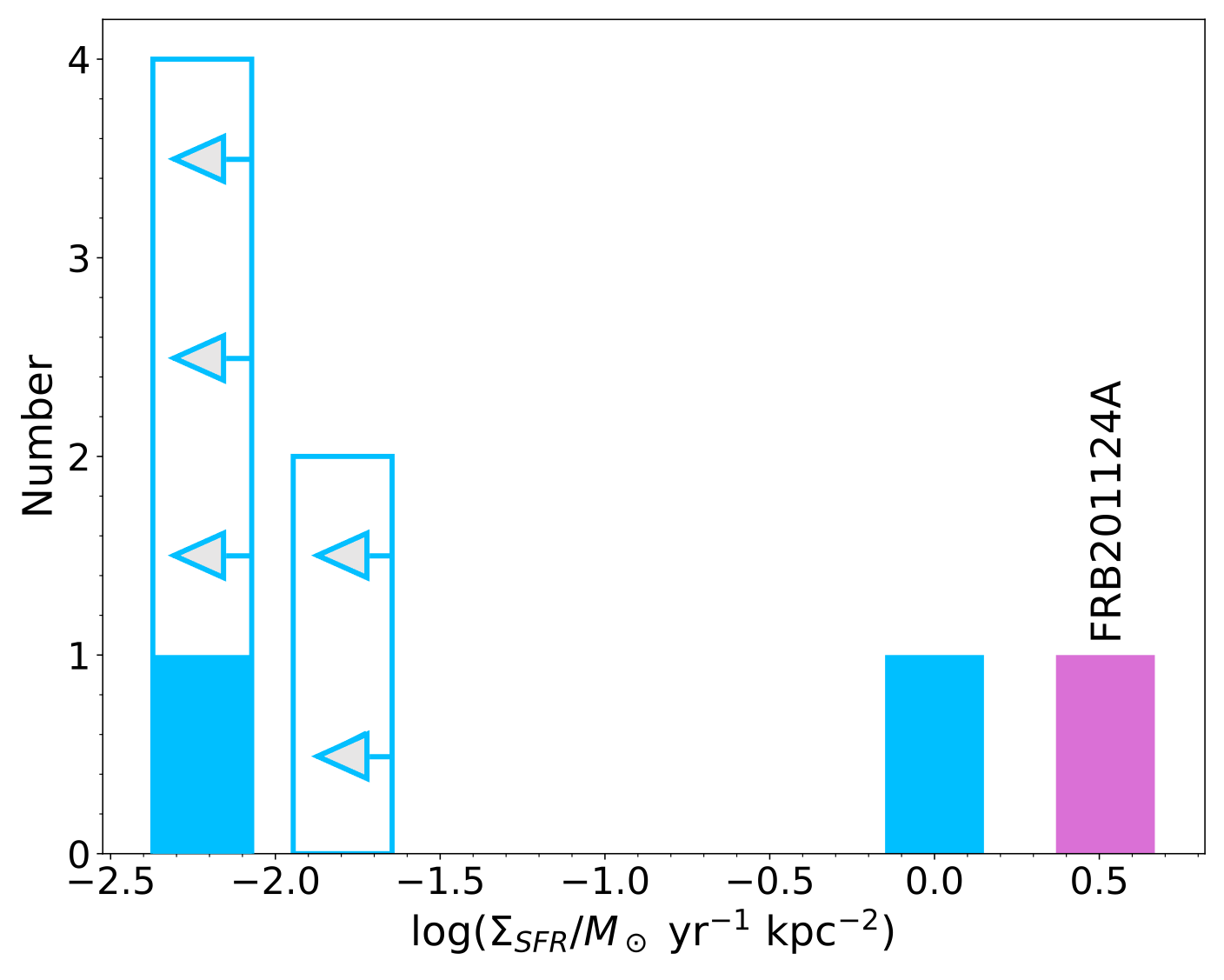}
\caption{Local star formation rate density $\Sigma_{SFR}(FRB)$ for \frb\ (this work; magenta bar)
compared to the sample of FRBs (from \citet{Mannings20} with the exception of \frba\ at 1 ${\rm \Msolar\ yr^{-1}\ kpc^{-2}}$, derived from  \citet{Bassa17}, see text). Arrows represent upper limits, and solid blue bars represent measurements  contained in each  bin width. 
}
    \label{fig:SigmaSFR}
\end{figure}

 The properties of the local environment and of the host  of  FRB provide a major clue on their progenitors. Different progenitor channels  are expected to yield distinct distributions of  location, local environment and host galaxy properties \citep[e.g.][]{Margalit19}.  
 In this regard, \frb\ exhibits some remarkable features.
   We find that it is located, with an accuracy of 360 pc, at the center of the
   most prominent region of star formation in the host galaxy, which is observed in the radio with  SFR$\approx 10\ {\rm M_\odot yr^{-1}}$.  This structure,  elongated by about 3 kpc and at a distance of  $0.8\pm0.4$ kpc from the center of the host galaxy embeds the FRB and might be associated to a spiral arm. 
The average  star formation surface density at the location of the FRB (Fig.\ref{fig:SigmaSFR}), $\Sigma_{SFR}(FRB)= 3.3\ {\rm M_\odot yr^{-1} kpc^{-2}}$, is two orders of magnitude larger than observed in all other FRB , with the  exception \frba\ that, with SFR$\approx 0.2-0.4\ {\rm M_{\odot}\ yr^{-1}}$ and a  size of 0.68 kpc \citep{Bassa17} has $\Sigma_{SFR}(FRB)\approx 1\ {\rm M_\odot yr^{-1} kpc^{-2}}$  \citep[see also][]{Mannings20} comparable to \frb. Such large surface star formation rates are  similar to those observed in Galactic star forming regions \citep{Evans09}. 
This region exhibits also an elevated local star formation rate in comparison to the mean global value of the host \footnote{Estimated by dividing the total SFR of the galaxy derived from H$\alpha$ with the area (at half light radius) of the galaxy, 19 kpc$^2$}  ($\frac{\Sigma_{SFR}(FRB20201124A)}{\Sigma_{SFR}(host)}\gtrsim 16$), while  in most  FRB the local SFR is comparable to the average of the host \citep{Mannings20}. 

The properties of \frb, particularly its location within a region of intense star formation,  favour a prompt  formation channel, with the magnetar formed
 from the SNe of a massive star.
Systematic studies of  host  properties \citep{Heintz2020, Bochenek21}, including the distribution of stellar masses and star formation rates,  find a wide range of host properties.  They  concluded that, although this variety can underline a diversity of progenitor channels, 
the observed global properties  are consistent with a   predominant population of young magnetars from core-collapse SNe. In this context, the global SFR and stellar mass derived for the host of \frb\ (Fig.\ref{fig:SFR})
 are in agreement  with a young magnetar formation  from a core collapse SNe.


With regard to the local environment,  this progenitor scenario would predict that FRBs should be preferentially located nearby to the brightest star forming regions of their host galaxies, as we observe in the case of \frb.  In contrast, HST observations of precisely localized FRBs  show that while FRB positions are consistent with spiral arms, they are not located on the most active star forming  regions \citep[][]{Bhandari20,Mannings20, Chittidi20}.
On a closer scale it is also expected that the positions  would be coincident or at small offset from star forming regions. This is due to the limited age (<10 kyr) of the active phase of the magnetar and its kick velocity after birth \cite[e.g.][]{Tendulkar21}.  However, two of the known repeaters, \frbb\ and \frba\ are found at an offset of $\approx 250$\ pc from the center of the closest star forming knots, a distance too large to be consistent with a young magnetar progenitor \citep{Bassa17,Tendulkar21}.
We find that \frb\ is consistent with being at the center of the SFR although, given  the error in the position, one cannot exclude an offset of $\approx 200$pc.  In such a case, an age of 0.5 Myr would be expected for a typical kick velocity of $\approx 400 {\rm km\ s^{-1}}$, questioning a young magnetar origin.
In conclusion, the properties of the environment of \frb, from  the host galaxy scale down to the local environment are consistent with those expected from a young magnetar progenitor from a SNe.


Considering the pieces of evidence discussed above, it is  difficult to draw any definitive conclusion on  FRB progenitors.
Different  channels for the formation of magnetar progenitors can  explain the broad distribution of properties of the host and local environment.  In this scenario \frb\ represents the clear example of a young magnetar progenitor from a core collapse SN.  The recent association of FRB\,20200120E with  a globular cluster indicates that at least another formation channel is present, with young magnetars produced  from accretion induced collapse of white dwarfs or   binary mergers of neutron stars \citep{Kirsten21}.
 Nonetheless,  the fact that the other well localized FRBs  are  close to star forming regions, suggests a predominant  channel associated with star formation \citep[e.g.][]{Ioka20,Tendulkar21} but with a broader range of SFR and offsets, in which \frb\ would then represent the  extreme in the distribution of local SFR. 
    \section{Limits to X-ray counterparts of FRBs}
    \label{sec: limitsmain}
    Notwithstanding the significant progress obtained via the study of the environment, and the increasing number of   precise locations that will be available in the future,  high resolution imaging of the local environment of FRBs provides only circumstantial evidence on the central source of FRBs. A key step forward may come from the identification of the counterpart of FRBs at other wavelengths.
    
Although our simultaneous  coverage of the \textit{Chandra} observation  with FAST and SRT observations does not find any significant association of  bursts from \frb  with X-ray photons  (Appendix \ref{sec:FRB}), it sets upper limits on the X-ray counterparts that can be confronted with a magnetar origin.
We place a 5$\sigma$ upper limit  ${\rm E_X\lesssim 6\times 10^{45}}$ erg in the 2-10 keV range on the X-ray energy of a burst
at the time of any radio burst and ${\rm E_X\lesssim 1.6\times 10^{46}}$ erg at any time during the \textit{Chandra} observation.
We  derive  a tighter upper limit ${\rm E_X\lesssim 1.1\times 10^{44}}$ erg assuming that X-ray bursts of similar fluence are emitted at the time of each radio  burst.
In Tab.\ref{tab:FRB} we compare the  limits with two other repeaters \frba\ \citep{Scholz17} and  \frbb\ \citep{Scholz20}.
The most constraining limit for X-ray bursts in coincidence with FRB is E$_X\lesssim 1.6\times 10^{45}$ erg for a single burst in  \frbb\ and ${\rm E_X\lesssim 1.1\times 10^{44}}$ erg in \frb, assuming that all FRB have an associated  X-ray burst.
 This corresponds to an upper limit on the ratio of the flare energy in  X-rays to radio $\rm{E_X/E_R\lesssim 5\times 10 ^{5}}$, and to a radio-to-X-ray spectral index  $\alpha_{RX}\gtrsim 0.4$ (F($\nu) \propto \nu^{-\alpha}$).   

X-ray bursts of known magnetars have isotropic energies much smaller than $10^{45}$ erg. The energy of the X-ray burst associated with  FRB-like events from  the Galactic magnetar SGR 1935+2154  is $\sim 10^{40}$ erg \citep{Li21}. Even for the most energetic giant flares detected from three known magnetars, the isotropic energies only reached a few times $10^{44}$ erg \citep{Woods06}. One may imagine more energetic X-ray flares for cosmological FRBs. If we assume that FRBs all have the X-ray-to-radio energy ratio as observed in  FRB-like events from the Galactic magnetar SGR 1935+2154, i.e. $ \rm{E_X/E_R= 3.6\times 10^4}$ \citep{Bochenek20}, the upper limit $\rm{E_X/E_R\lesssim 5\times 10 ^{5}}$ derived from our observations is still consistent with such an assumption. 
We therefore draw the conclusion that the magnetar flaring origin of FRBs for this source cannot be disfavored and is consistent with the available data. 

    \section {Conclusions}
    \label{sec: concmain}
    In this paper we have presented the results of a campaign involving nine facilities from radio to X-rays, and aimed at constraining the multiwavelength properties of the quiescent and flaring emission of \frb, the  third closest repeater recently localized.

We find that \frb\ is located at the center of an extended radio source associated to the brightest region of star formation in the host galaxy, corresponding to  SFR$\approx 10\, M_\odot yr^{-1}$. 
This structure is elongated by about 3 kpc and at a distance of $0.8\pm0.4$ kpc from the center of the host galaxy. 
The  star formation surface density at the location of the FRB (Fig.\ref{fig:SigmaSFR}), $\Sigma_{SFR}(FRB)= 3.3\ {\rm M_\odot yr^{-1} kpc^{-2}}$, is two orders of magnitude larger than typically observed in  FRB \citep{Mannings20}, 
but similar to that observed in Galactic star forming regions
\citep{Evans09}. 
This region exhibits also an elevated local star formation rate in comparison to the mean global value of the host ($\frac{\Sigma_{SFR}(FRB20201124A)}{\Sigma_{SFR}(host)}\gtrsim 16$), while in most  FRB the local SFR is comparable to the average of the host. 
All the properties of the environment of \frb, from  the host galaxy scale down to the local environment are  consistent with a young magnetar progenitor  from a core collapse SNe.
In contrast, high resolution observations of the local environment  of other precisely located FRBs does not support unambiguously this association \citep{Mannings20, Tendulkar21}, suggesting a diversity of progenitors.

A search  for X-ray counterparts of FRB  detected by FAST and SRT,
carried out with \textit{Chandra},  sets upper limits on the X-ray counterparts that can be confronted with a magnetar origin.
The tightest upper limit ${\rm E_X\lesssim 1.1\times 10^{44}}$ erg is derived assuming that X-ray bursts of similar fluence are emitted at the time of each radio flare.
 This corresponds to an upper limit on the ratio of the flare energy in  X-rays to radio $\rm{E_X/E_R\lesssim 5\times 10 ^{5}}$, 
consistent with that observed in  FRB-like events from the Galactic magnetar SGR 1935+2154 \citep{Bochenek20}.


 Future observations with higher spatial resolution in the optical and radio bands, with adequate sensitivity in radio to low surface brightness,  should clarify the morphology of the $\approx 3$kpc star forming region observed in radio, its association with host galaxy structures such as spiral arms, and  further resolve  its fine structure, so to test the young magnetar scenario with a finer measurement of the offset between the FRB and the SFR region. 



\begin{acknowledgements}
We acknowledge support from a grant PRIN-INAF SKA-CTA 2016. The National Radio Astronomy Observatory is a facility of the National Science Foundation operated under cooperative agreement by Associated Universities, Inc. We thank the staff of the GMRT that made these observations possible. GMRT is run by the National Centre for Radio Astrophysics of the Tata Institute of Fundamental Research. The European VLBI Network is a joint facility of independent European, African, Asian, and North American radio astronomy institutes. Scientific results from data presented in this publication are derived from the following EVN project code(s): RP032A. e-MERLIN is a National Facility operated by the University of Manchester at Jodrell Bank Observatory on behalf of STFC.
We thank the {\it Swift} and {\it Chandra} staff for the support in carrying out the ToO observations. The Sardinia Radio Telescope (SRT) is funded by the Department of University and Research (MIUR), the Italian Space Agency (ASI), and the Autonomous Region of Sardinia (RAS) and is operated as National Facility by the National Institute for Astrophysics (INAF).  We thank an anonymous referee for useful comments.
\end{acknowledgements}



\bibliographystyle{aa}
\bibliography{gwgrb_references} 


\begin{appendix}

\section{VLA discovery of a quiescent radio source}
\label{sec:VLA}
We pointed the FRB position from \cite{2021ATel14515....1D} with the VLA on April 9, May 11, and June 22, 2021 (project code SG9112). The first two runs were performed in D-configuration, while the latter in C. The observing bands were S (2-4 GHz) and X (8-12 GHz) for the first epoch, S, C (4-8 GHz), and X for the second one, and L (1-2 GHz), X, and K (18-26 GHz) for the third one (see Tab. \ref{tab:radio} for a summary). The calibrator 3C\,147 was used both for amplitude scale and bandpass corrections, while the phase calibrator paired to the target was J0534+1927 for the first two epochs ($\sim9^\circ$ apart), and J0431+2037 for the third one ($\sim10^\circ$ apart), due to the configuration change. Data were reduced with the Common Astronomy Software Applications package \citep[CASA;][]{2007ASPC..376..127M}. We processed the raw data with pipeline version 5.6.2. The calibrated visibilities were then imaged with the task {\tt{TCLEAN}}, applying natural weighting to optimize sensitivity. 

A constant radio source was detected in the different epochs, offset by $\sim$3.3 arcsec with respect to the \cite{2021ATel14515....1D} localization, but consistent with later ones \citep{Ricci21}. The flux density was consistent within errors in the different epochs, at the common frequencies of 3, 9, and 11 GHz (see Tab. \ref{tab:radio}). 
The estimated radio luminosity is $L_{\rm{5\,GHz}}=3\times10^{38}\,\rm{erg\,s^{-1}}$. 
The morphology is unresolved at angular resolutions larger than $\sim$1\arcsec, while the 22 GHz observations in C-configuration from the latest epoch revealed an elongated emission region, with an extension of $\sim$2\arcsec. A Gaussian fit with a single component results in a deconvolved size of 2.2$\pm$0.6$\times$0.5$\pm$0.3\arcsec, with a position angle of 140$\pm$8$^\circ$, confirming the resolved morphology. The component is centered at RA 05:08:03.50$\pm$0.01s, Dec +26:03:38.36$\pm$0.18\arcsec, consistent with the  position of the  FRB.  The astrometric accuracy was estimated by adopting a conservative VLA positional uncertainty of 10\% of the FWHM\footnote{\url{https://science.nrao.edu/facilities/vla/docs/manuals/oss/performance/positional-accuracy}} (0.11\arcsec), and summing that in quadrature with the peak location error from the Gaussian fit. We obtained a total uncertainty for the Gaussian centroid location  of $\pm$0.21\arcsec.


\begin{table*}
\centering
 	\caption{Journal of radio observations of the quiescent source associated with \frb.}
 	\label{tab:radio}
 	\begin{tabular}{ccccccccccc}
    \hline
Telescope  &  Date          & Frequency & Bandwidth  & Time on source    & Flux density  &       FWHM             \\
           & (dd/mm/yyyy)    &   (GHz)   &  (GHz)    & (min)             &  ($\mu$Jy)    &       (\arcsec)        \\
\hline
uGMRT       & 24/07/2021    &  0.19     &  0.125     & 45                & $<$3600       &       21.6$\times$10.4 \\
uGMRT       & 21/07/2021    &  0.38     & 0.24       & 98                & 1700$\pm$200  &       7.0$\times$4.9   \\
VLA-D       & 09/04/2021    &  3        &  2         & 28                &  340$\pm$30   &       18.0$\times$16.1 \\
            &               &  9        &  2         & 33                &  150$\pm$10   &       7.4$\times$6.7   \\
VLA-D       & 11/05/2021    &  3        &  2         & 38                &  335$\pm$18   &       31.7$\times$16.9 \\
            &               &  5        &  2         & 11                &  259$\pm$10   &       20.2$\times$11.4 \\
            &               &  7        &  2         & 11                &  165$\pm$11   &       18.5$\times$8.2  \\
            &               &  9        &  2         & 22                &  159$\pm$7    &       10.4$\times$6.8  \\
            &               &  11       &  2         & 22                &  126$\pm$10   &       8.4$\times$5.4   \\
VLA-C       & 22/06/2021    &  1.5      &  1         & 14                &  706$\pm$76   &      12.8$\times$12.1  \\
            &               &  9        &  2         & 10                &  142$\pm$12   &       2.6$\times$2.6   \\
            &               &  11       &  2         & 10                &  104$\pm$14   &       2.3$\times$2.1   \\
            &               &  22       &  8         & 26                &   78$\pm$8    &       1.1$\times$1.0   \\
\hline 
e-MERLIN    & 11/07/2021    &  5        & 0.5        & 500               &  $<$85        &       0.085$\times$0.045     \\
EVN         & 12/05/2021    &  5        & 0.1-0.5    & 245               &  $<$54        &       0.0035$\times$0.0017   \\
\hline  
    \end{tabular}
\end{table*}



\section{ X-ray imaging with \textit{Swift} and \textit{Chandra}}
\label{sec:Chandraimage}

We carried out observations (PI: Piro) with \textit{Swift}/XRT in PC mode starting at 2021-04-06 19:47:02 UT and ending at 2021-04-07 14:53:17 UT for a total of 9.9 ks exposure under ObsIDs 00014258001 and 00014258002.
The data were processed using the \texttt{xrtpipeline} task, and the individual ObsIDs were stacked using \texttt{XSELECT}. Results are presented in \S \ref{sec: xraymain}.

\textit{Chandra X-ray Observatory} (CXO) observations (ObsID: 25016; PI: Piro), carried out under Director's Discretionary Time (DDT), occurred on April 20, 2021 for a total of 29.7 ks. 
Data were processed using the \textit{Chandra} Interactive Analysis of Observations (\texttt{CIAO} v.~4.13; Fruscione et al. 2006) software and the latest calibration database (\texttt{CALDB} v.~4.9.4).
Our analysis was performed in the $0.5-7$ keV energy range.
The native astrometry of the image was corrected by aligning 2 common point sources in the Sloan Digital Sky Survey (SDSS) Catalog. We restricted these sources to those within $2.5$ arcmin of the \textit{Chandra}/ACIS-S3 detector's optical axis. This resulted in an astrometric shift by $0.5$\arcsec, with tie uncertainty $\sim0.25$\arcsec. Further analysis of the astrometric uncertainty using a total of 4 common point sources with SDSS, including those at $\gtrsim 2.5$ arcmin from the optical axis of ACIS-S3, yielded a consistent astrometric solution and did not change the result.

A weak X-ray detection at the position of the EVN VLBI \citep{Marcote21} with a
0.5-7.0 keV count-rate of 1.0$^{+0.7}_{-0.6}$ $\times$ 10$^{-4}$ cts s$^{-1}$ 
corresponds to a luminosity of 
$\rm{L_X=(3.1 \pm 1.9 )\ 10^{40}}$ erg  s$^{-1}$, 
assuming that absorption is limited to our Galaxy (\S\ref{sec: xraymain}).
If we also take into account the large DM=$421\pm4$ associated to the FRB (Appendix \ref{sec:FRB}), 
the intrinsic absorption component could be as high as $N_{H,z} \approx 1.2\times 10^{22}$ cm$^{-2}$ \citep{He13}. This would imply an unabsorbed flux $\rm{F_X \approx 1.8 \times 10^{-15}}$ erg cm$^{-2}$ s$^{-1}$ (2-10~keV) and a luminosity  
$\rm{L_X \approx 4 \times 10^{40} erg \ s^{-1}}$.

These values fit well within the correlation between FIR and X-ray luminosities in star-forming galaxies \citep{Ranalli2003}, and imply a recent star formation rate SFR = 10$^{+11}_{-7}$ \,$M_{\odot}$~yr$^{-1}$ \citep{Lehmer2016}. 
Whereas radio and FIR indicators probe the SFR over $\approx$100~Myr timescales, 
studies of local star-forming galaxies show that the dominant component 
of the X-ray emission are high-mass X-ray binaries (HMXBs) associated with a young ($\lesssim$30 Myr) stellar population \citep{Grimm2003}. 





\section{Follow-up of the persistent radio source with EVN, e-MERLIN, and uGMRT}
\label{sec:EVN}

Following the detection of a persistent radio source with the VLA, we triggered a Very Long Baseline Interferometry (VLBI) campaign involving the European VLBI Network (EVN) and the enhanced Multi Element Remotely Linked Interferometer Network (e-MERLIN). EVN observations were performed on May 12th, 2021, in e-VLBI mode with the standard setup at 5 GHz (project code RP032A, PI: Panessa). The array was composed by 9 EVN antennas (IR, BD, SV, HH, SH, O8, EF, WB, JB) plus 6 from e-MERLIN (JL, PI, KN, DE, DA, CM). The data rate was 2 Gbps for all antennas except for SV and BD (1 Gbps), and the e-Merlin array (512 Mbps). This reflected into different bandwidths, from 0.1 to 0.5 GHz. Sources J0237+2848 and J0555+3948 were used as fringe finders at the beginning/end of the experiment, while J0502+2516 for target phase-referencing (1.4$^\circ$ apart). A total on-source time of 4 hours was reached. Data were reduced with the Astronomical Image Processing System (AIPS) software, following standard calibration procedures for continuum data. Visibilities were then imaged using the DIFMAP software \citep{1997ASPC..125...77S}, applying natural weighting. The obtained angular resolution was 3.5$\times$1.7 mas, with a position angle of 7.15$^\circ$. The RMS was 9 $\mu$Jy/beam. 

We did not detect any compact source above a confidence level of 6 sigma, resulting in an upper limit of 54 uJy/beam. This result confirms with deeper observations the previous one from the PRECISE project \citep{Marcote21}, also obtained with the EVN but at 1.4 GHz. Those authors concluded that the nature of the emission detected with the VLA must be of extended nature, not being visible at VLBI resolution. From non-detections on the shortest baseline of the array (Effelsberg-Westerbork; $\sim$270 km) they could estimate an angular size $\gtrsim$140 mas for the extended emission, corresponding to a projected linear size $\gtrsim$260 pc. 
In our observation, the sub-array composed by e-MERLIN antennas plus JB - i.e. the array baselines more sensitive to extended emission - gave an angular resolution of 60$\times$36 mas, and an RMS of 43 $\mu$Jy/beam (applying natural weighting). Considering a threshold of 6 sigma, the e-MERLIN non-detection implies a surface brightness $<$0.045 Jy/kpc$^2$. 

Given the results of the EVN run, we performed deeper observations with e-MERLIN at 5 GHz on July 11, 2021 (DDT project DD11007, PI: Bruni), with the aim of recovering the SFR emission at an intermediate scale between our previous VLA and EVN observations. The run originally included also 1.5 GHz, the frequency band where most of the flux density was expected, however the sensitivity was hampered by the lack of suitable, unresolved, phase calibrators near to target. At 5 GHz, we observed with the full 512 MHz bandwidth, for a total of $\sim$11 hours on target, reaching an RMS of 17 $\mu$Jy/beam. The phase calibrator was J0506+2141, with an angular separation of 4.4$^\circ$. The angular resolution, applying natural weighting, was 85$\times$45 mas. Also at this sensitivity, no emission was detected at a significance level larger than 5 sigma (see Fig. \ref{fig:image}),
implying a surface brightness $<$0.008 Jy/kpc$^2$.

To further complete the frequency coverage on the persistent radio source, we requested DDT observations with the upgraded Giant Metrewave Radio Telescope (uGMRT). Observations were carried out on July 21 and 24, 2021 (project code ddtC194, PI: Bruni), using the band-3 (260-500 MHz) and band-2 (125-250 MHz) receivers, respectively. The amplitude scale and bandpass calibrations were performed on 3C\,147, while the adopted phase calibrator was J0534+1927. Data were reduced with the {\tt{CAPTURE}} pipeline \citep{2021ExA....51...95K}. In band-3, the final image RMS was 50 $\mu$Jy/beam, and the angular resolution 7.0\arcsec$\times$4.9\arcsec. We detected the source as an unresolved component with a flux density of 1.7$\pm$0.2 mJy. Band-2 was hampered by radio frequency interferences, with more than 50\% of data flagged on each antenna. The final image showed no detection at an RMS level of 1.2 mJy/beam, resulting in a 3$\sigma$ upper limit of 3.6 mJy. The angular resolution was 21.6\arcsec$\times$10.4\arcsec.


\section{The radio spectrum from star formation in the host galaxy}
\label{sec:radiospectrum}
Star-forming galaxies emit both thermal (free-free) and non-thermal (synchrotron) radiation in the radio regime. The synchrotron component, resulting from SNR, dominates at frequencies up to about  20 GHz, while free-free emission becomes significant at higher frequencies \citep{Klein18}. At low frequencies, typically   well below 1 GHz, free-free absorption takes place \citep{Schober17}.
The resulting spectrum is thus described as

\begin{equation}
F_{\nu}=C_S 
[\frac{\nu}{GHz}]^{-\alpha_s}
 e^{-\tau_{ff}}+C_{ff}[\frac{\nu}{GHz}]^{2}(1-e^{-\tau_{ff}})
\end{equation}
with 
\begin{equation}
\tau_{ff}(\nu)=[\frac{\nu}{\nu_{ff,abs}}]^{-2.1}
\end{equation}

 The first term in the equation is the synchrotron component, modelled with a power law with a slope of $\alpha_s$.
 The second, flatter component, $\propto\nu^{-0.1}$ in the optically thin regime, is due to free-free  emission. In fig.\ref{fig:broadband_SED_wRadio} we present the best fit model on our VLA and GMRT data (Tab.\ref{tab:radio}).
 The radio spectrum is dominated by the synchrotron component with $\alpha_s=0.76\pm0.07$, no evidence of absorption and an upper limit (3$\sigma$) of $\lesssim 10\%$ of the thermal fraction  at 1.4 GHz. 
Those values are consistent with the range  observed in  radio  spectra of star forming galaxies \citep{Tabatabaei17}.


\section{Optical spectroscopy}
\label{sec: opt_spectroscopy}

We obtained optical spectroscopy of the putative host galaxy of FRB 20201124A on 2021-04-21 starting at 03:18:35.51 with the DeVeny spectrograph mounted on the 4.3-m LDT for a total exposure of 4$\times$600 s. DeVeny was configured with the 300 g mm$^{-1}$ grating and a 1.5\arcsec slit width. The slit was aligned at position angle $63^\circ$ East of North and covers the location of the FRB determined by the EVN VLBI localization. However we note that the width of the slit does not cover completely either the full extension of persistent radio source or the host galaxy.
We reduced the data using standard procedures in the IRAF package to perform bias subtraction, flat field correction, and cosmic ray removal with \texttt{L.A.Cosmic} \citep{vanDokkum2001}.

\begin{figure}
 \includegraphics[width=\columnwidth]{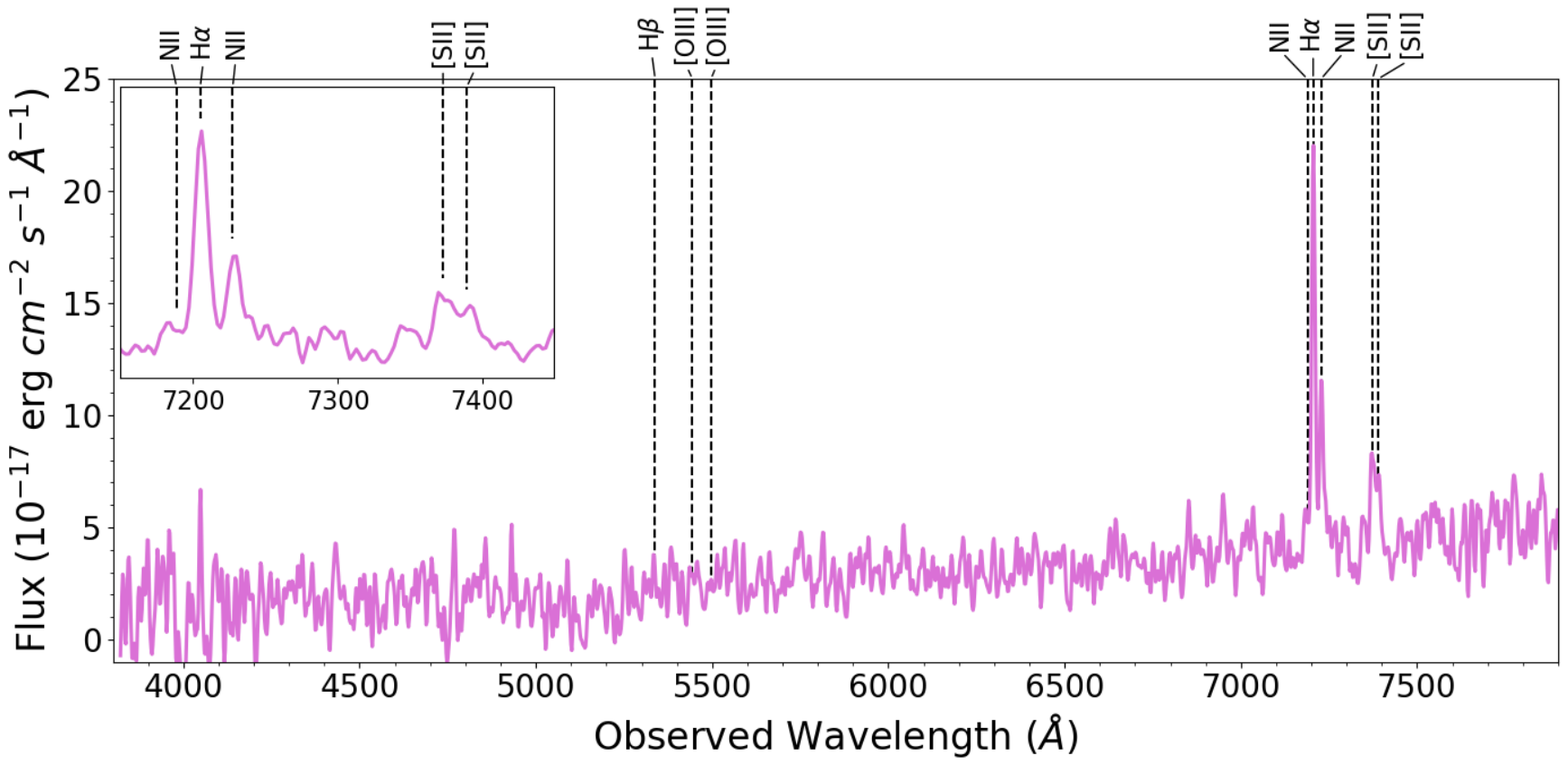}
\caption{
LDT/DeVeny optical spectrum of the host galaxy of \frb\ at $z=0.0978 \pm 0.0002$. The spectrum (solid purple line) has been smoothed with a Gaussian kernel. Dashed black lines indicate the expected location of emission lines at this redshift; we do not observe Oxygen lines or H$\beta$. The inset figure shows a zoom on the region of the detected emission lines between $7150$ and $7450\,\AA$.
The spectrum is not corrected for Galactic extinction.}
    \label{fig:host_spectrum}
\end{figure}

The spectrum was calibrated using the spectrophotometric standard star Feige 34. The resulting spectrum is displayed in Figure \ref{fig:host_spectrum}.
We detect emission features at $\lambda_\textrm{obs}\approx 7207$, $7230$, $7374$, and $7391\,\AA$ associated to H$\alpha$, the [NII] doublet, and the [SII] doublet. 
Given the sensitivity of these observations, our non-detection of H$\beta$ or Oxygen lines is consistent with the fluxes reported by \citet{Fong21}. 
Using the observed lines, we derive a redshift $z=0.0978 \pm 0.0002$. 
Line properties were derived by fitting the lines with Gaussian functions using the \texttt{specutils} package in \texttt{Python}. The emission line fluxes were corrected for Galactic extinction $E(B-V)=0.63$ mag \citep{Schlafly2011} assuming a \citet{Cardelli1989} extinction law. The line properties are reported in Table \ref{tab:emlines}.

From the H$\alpha$ emission line, we derived an SFR$_{\textrm{H}\alpha}=0.89\pm0.16$ $M_\odot$ yr$^{-1}$ \citep{Kennicutt1988}, assuming a Chabrier  initial mass function  \citep[IMF,][]{Chabrier2003}.
From the analysis of the host galaxy's spectral energy distribution (SED; Appendix \ref{sec: Galaxy SED modeling}) we inferred a global value of $A_V\sim1.3$ mag for the intrinsic extinction. By applying this correction to our spectrum, the optically-derived SFR is
 SFR$_{\textrm{H}\alpha}=2.3\pm0.4$ $M_\odot$ yr$^{-1}$.

\begin{table}
\centering
\caption{Host galaxy emission line properties from our optical spectroscopy with LDT/DeVeny. These values are corrected for Galactic extinction $E(B-V)=0.63$ mag \citep{Schlafly2011}.}
\label{tab:emlines}
\begin{tabular}{lcc}
Line & $\lambda_\textrm{obs}$ & Flux \\
 \hline
&[$\AA$] & [$10^{-15}$ erg cm$^{-2}$ s$^{-1}$]\\
 \hline
 H$\beta$ & 5345.1 & $<2.9$ \\
H$\alpha$ & 7206.8 & $7.4\pm1.3$\\
 NII$_{\lambda6585}$ & 7229.6 & $3.9\pm1.1$  \\ 
 SII$_{\lambda6718}$ & 7373.5 & $2.9\pm0.7$ \\
SII$_{\lambda6732}$ & 7391.3 & $2.2\pm0.7$ \\
 \hline
\end{tabular}
\end{table}


\section{Galaxy SED analysis}
\label{sec: Galaxy SED modeling}

We modeled the spectral energy distribution (SED) of the host galaxy using \textsc{prospector} \citep{Johnson2019} and the methodology outlined in \citet{OConnor2021}. 
We made use of archival photometry from SDSS, the Two Micron All Sky Survey \citep[2MASS;][]{Skrutskie2006}, and the Wide-field Infrared Survey Explorer \citep[WISE;][]{Cutri2021}. This photometry covers optical and near-infrared wavelengths $ugrizJHK$ in addition to two WISE infrared bands ($W1$ and $W2$). We exclude the WISE $W3$ and $W4$ photometry due to the uncertainty surrounding thermal dust emission models \citep[see, e.g.,][]{Leja2017}. 
From the archival SDSS/$u$-band image, we derive a $3\sigma$ upper limit $u\gtrsim 22.0$ mag on an underlying source. This is consistent with the deeper limit from \textit{Swift}/UVOT ($u\gtrsim 22.8$ mag), which we include in our modeling.
We also include the measured H$\alpha$ emission line flux as pseudo narrow-band photometry in order to better constrain the SFR. The photometry was corrected for Galactic extinction in the direction of the FRB, $E(B-V)=0.63$ mag \citep{Schlafly2011}, prior to modeling with \textsc{prospector}.
We adopt a \citet{Chabrier2003} IMF with integration limits of 0.08 and 120 $M_{\odot}$ (\texttt{imf\_type = 1}), an intrinsic dust attenuation using the extinction law of \citet{Cardelli1989} and accounting for both additional dust in  nebular regions (\texttt{dust1}) and diffuse dust throughout the galaxy (\texttt{dust1}), and a  delayed-$\tau$ (\texttt{sfh=4}) star-formation history characterized by an e-folding timescale, $\tau$. We account for the contribution of nebular emission using the photoionization code \textsc{Cloudy} \citep{Ferland2013}. The synthetic spectral energy distributions (SEDs) corresponding to these models were computed with the flexible stellar population synthesis (FSPS) code \citep{Conroy2009}. The free parameters in these models are the total stellar mass formed $M$, the age $t_{\rm age}$ of the galaxy, the e-folding timescale $\tau$, the intrinsic reddening $A_V$, and the metallicity $Z_*$. We adopt uniform priors in log $t_{\rm age}$, log $\tau$, log $Z$, $E(B-V)$ as in \citet[][cf. their Table 2]{Mendel2014}. From these parameters, we derive the stellar mass, $M_*$, the mass-weighted stellar age, $t_m$, and the star-formation rate, SFR. 

\begin{figure}
\includegraphics[width=1.0\columnwidth]{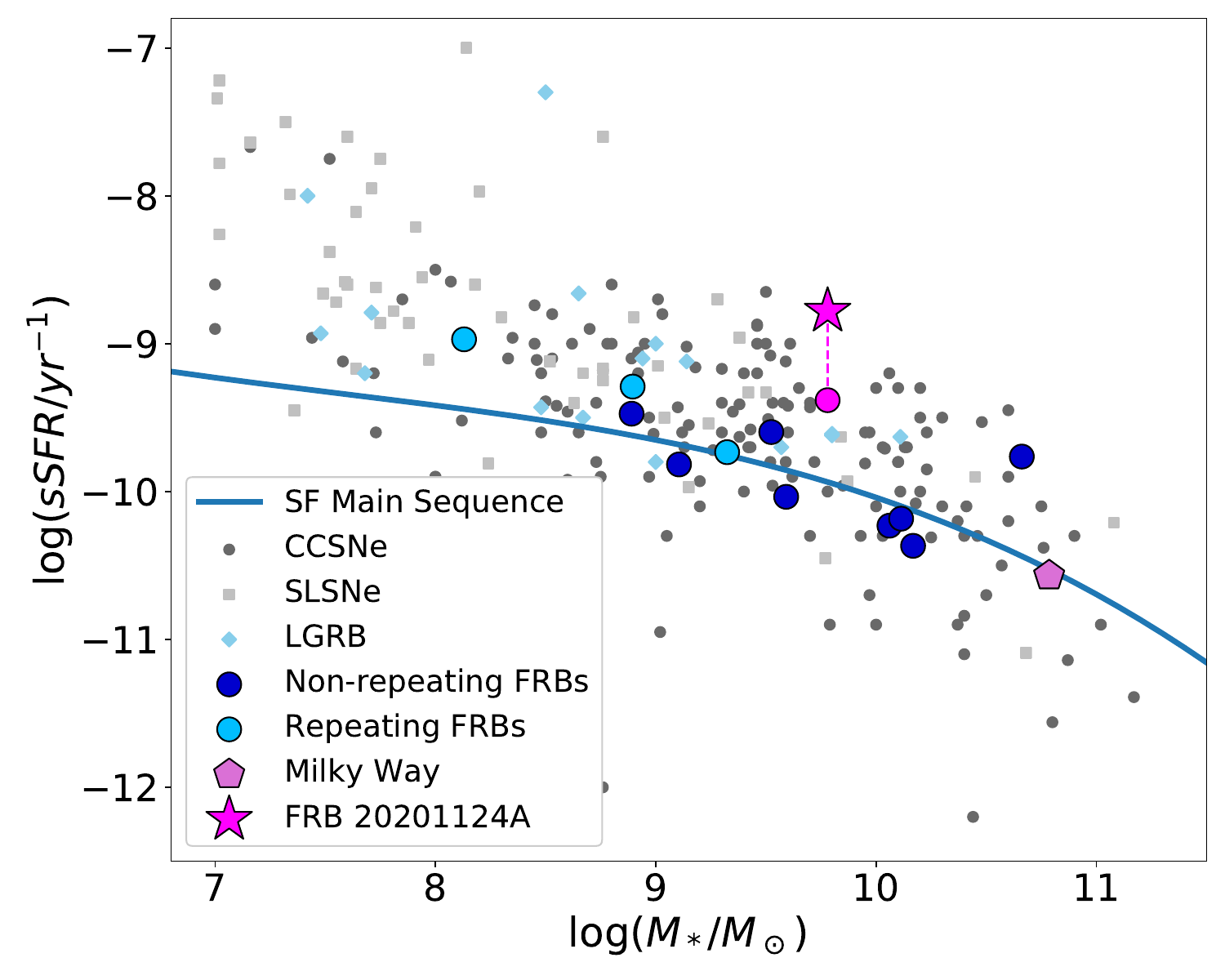}	
	
    \caption{ 
    Specific star formation (SFR normalized to the total mass of the galaxy) of \frb\ (magenta star), other repeating FRBs (light blue circles), non-repeating FRBs (dark blue circles) from \citet{Heintz2020} and our own Galaxy \citep[light purple pentagon;][]{Licquia2015}. For \frb\ we display the sSFR derived from radio observations (Appendix \ref{sec:VLA}) with a  magenta star, and a dashed line connects this value to the magenta circle representing the sSFR derived from optical spectroscopy (Appendix \ref{sec: opt_spectroscopy} and Fig.\ref{fig:host_spectrum}).
    We also display the populations of other astrophysical transients for reference by including the low redshift ($z<0.3$) populations of long Gamma-ray Bursts (LGRBs; small blue diamonds), core-collapse supernovae (CCSNe; small gray circles), and superluminous supernovae (SLSNe; small gray squares) from \citet{Taggart2021}. These values are compared to the low redshift ($z<0.05$) star forming main sequence from \citet{Saintonge2016} displayed by a solid line. Our object is located well above the star formation main sequence, above which galaxies have enhanced star formation efficiencies.
    }
    \label{fig:SFR}
\end{figure}

\begin{figure}
 \includegraphics[width=\columnwidth]{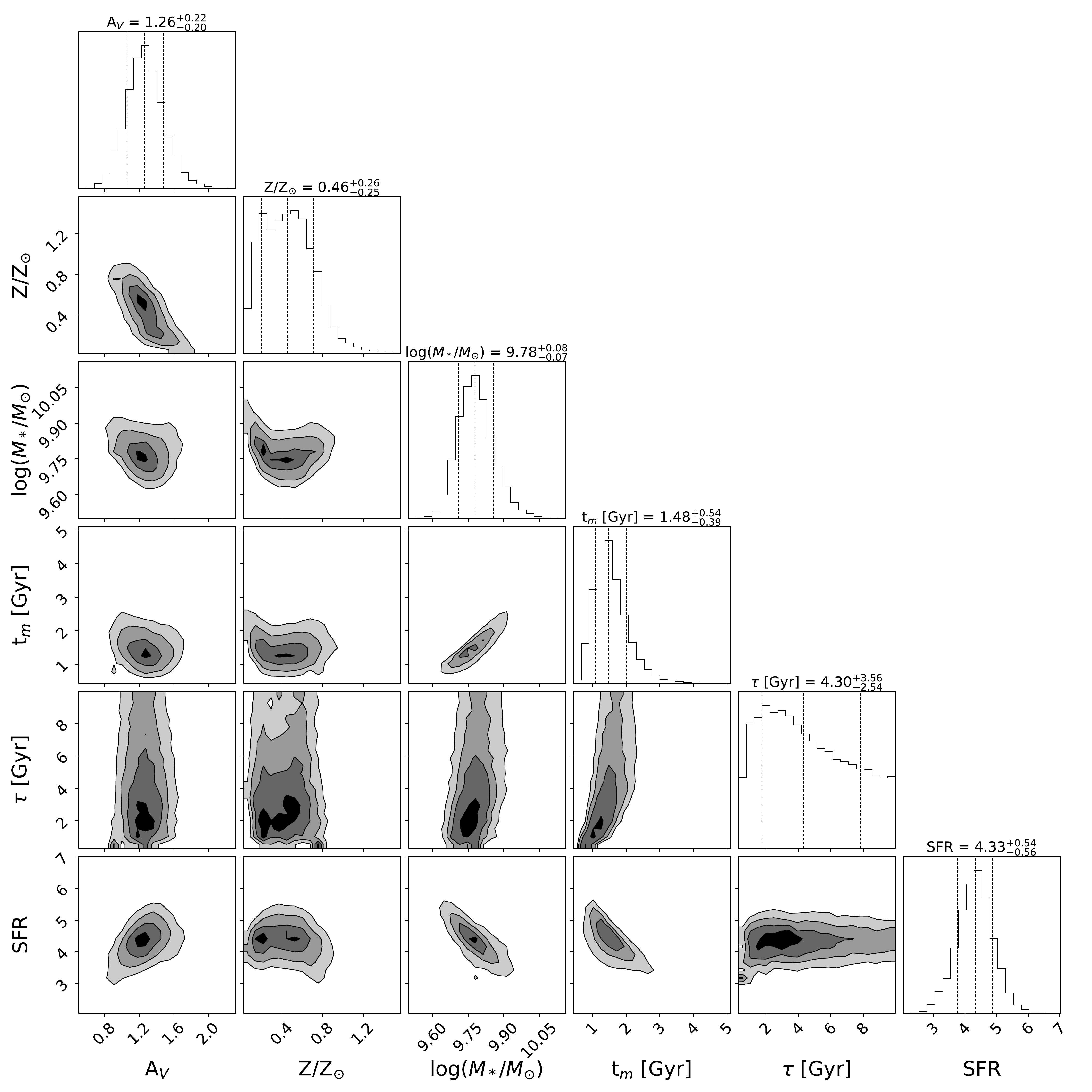}
\caption{ Corner plot demonstrating the results of our \texttt{prospector} modeling of the host galaxy SED as outlined in Appendix \ref{sec: Galaxy SED modeling}.
}
\label{fig:corner_plot}
\end{figure}

The best fit model spectrum is shown in Figure \ref{fig:broadband_SED_wRadio}, the corner plot in Figure \ref{fig:corner_plot} and the best fit parameters reported in Sect.\ref{sec: sfrmain}. 
We note that the mass-weighted stellar age and stellar mass are slightly smaller compared to those presented by \citet{Ravi21} and \citet{Fong21}. We find that this is likely due to different model assumptions within \texttt{prospector}.
We compare the stellar mass and star formation rate to the host galaxies of CCSNe, LGRBs, and other FRBs in Fig. \ref{fig:SFR}.

The  X-ray luminosity expected from such a galaxy  is derived as follows. There are three main components of galaxy X-ray emission: low-mass X-ray binaries (LMXB), high-mass X-ray binaries (HMXB), and hot gas. The strengths of these components can be modeled as a function of galaxy properties such as $M_*$ and SFR. We adopt the recipe from \citet{Fragos2013} and \citet{Yang2020}:
\begin{equation}
   \log(L_{2-10 keV}^{LXRB}/M_*)=f(t_m),
   \end{equation}
   \begin{equation}
    \log(L_{2-10 keV}^{HXRB}/SFR)=g(Z_*),
       \end{equation}
       \begin{equation}
    \log(L_{0.5-2 keV}^{hot}/SFR)=38.9 
\end{equation}
where luminosities are in erg s$^{-1}$. The functions   $f(t_m)$ and $g(Z_*)$ are given in \citet{Fragos2013}. Substituting the best fit values one obtains $L_{(2-10 keV)}=(5\pm 2)\times 10^{40}$ erg s$^{-1}$,  consistent with the observed X-ray luminosity. This emission is dominated (80\%) by the HMXB component associated with young stellar populations.

\section{Origin of the persistent emission from radio to X-rays} 
\label{sec:persistentorigin}
Our EVN observation demonstrated that the emission is extended and that the bulk of the
radio emission must arise on scales larger than about 200 pc. 
We argue that this emission  is  produced by  star formation in the host galaxy. First, the radio luminosity in galaxies is correlated  with SFR \citep{Murphy11} as SFR= $\frac{L_{1.4GHz}}{1.7\times 10^{28} {\rm erg\ s^{-1} Hz^{-1}}} {\rm M_\odot yr^{-1}}$ that, in the case  of \frb\ gives SFR$\approx 10 M_\odot yr^{-1}$.  This is consistent with the value derived from X-rays, and also in agreement with the smaller values derived from from optical/IR (2-4 $M_\odot yr^{-1}$), considering that the radio  may include additional emission from  heavily  obscured star formation \citep{Prescott07,Dye08} or from enhanced star formation that may have occurred over the longer time scale ( $\lesssim 100$ Myr) probed by radio observations \citep{Condon92}. Furthermore, the radio spectrum in the 0.4-10 GHz range has a slope of $0.76\pm0.07$, that is consistent with the average of $0.7\pm 0.4$  observed in star forming galaxies \citep{Ibar2009}.

 Alternative scenario involving a low-surface-brightness AGN-driven radio emission, e.g. jets, are unlikely \citep{Biggs10}. 
 A low-luminosity radio quiet AGN, like low-ionization nuclear emission-line regions and Seyfert galaxies    would be characterized by a more compact morphology
 \citep[e.g.,][]{Panessa2013}.
 A small fraction of radio active nuclei are in the so-called dying or remnant phase. The active phase of a radio AGN can last several tens of Myr, after which the nuclear activity stops and the source starts to fade away \citep[e.g.,][]{Parma1999}. The core and the jets disappear and only the lobes remain visible, radiating their energy away. Such a scenario can account for the non-detection of the compact core component. However, the observed properties of the extended radio source in \frb\ are markedly different from those observed in the remnant phase of radio galaxies. First, the radio spectrum of the remnant of Radio Loud (RL) AGN is much brighter at lower frequency, with a very steep spectrum ($\alpha \gtrsim 1$ above a few hundred MHz), due to the exponential cut-off from rapidly cooling electrons. Furthermore, the physical size of the region and the total radio power associated with remnants are respectively about two and three orders of magnitude larger than observed in our case, excluding this scenario.

\section{Chandra  limits on X-ray counterparts of FRB detected by FAST and SRT}
\label{sec:FRB}

We have organized a simultaneous coverage of the \textit{Chandra} observation  with FAST and SRT observations aimed at searching for X-ray counterparts of FRB.
\textit{Chandra} observation took place  on April 20, from 5:28 to 13:50 UT  and lasted for 8.3 hrs. 

 SRT \citep{bolli,prandoni} observed \frb\ on 20 April
2021 for 3 hours starting at 10:30 UT. Observations were performed both at 1.5 GHz and 330 MHz.
Data at 1.5~GHz were recorded with the ATNF digital backend
PDFB3\footnote{\url{http://www.srt.inaf.it/media/uploads/astronomers/dfb.pdf}} in search mode
over an effective bandwidth of 420\,MHz split into 1\,MHz channels. Total intensity data were 2-bit
sampled every 100\,$\mu$s. Data at 330~MHz were recorded with the ROACH1 digital backend in
baseband mode over a bandwidth of 64\,MHz which was subsequently split into 0.25\,MHz
channels. The raw data were coherently dedispersed  (removing the intrachannel smearing only) at the DM of the FRB (DM = 413\,pc\,cm$^{-
3}$,  as quoted by, i.e., \citealt{ATel14497}) and the resulting total intensity data were 8-bit sampled every 128\,$\mu$s.
A search for single pulses was performed on the data using the {\sc spandak}
pipeline\footnote{\url{https://github.com/gajjarv/PulsarSearch}} \citep{gajjar18}.
The pipeline uses {\sc rfifind} from the {\sc presto}
package\footnote{\url{https://github.com/scottransom/presto}} \citep{ransom2001} for high-level radio
frequency interference (RFI) excision.
 It then performs a first search for bursts through {\sc Heimdall} \citep{bbb+12}. A DM range from 0 to 1000\,pc\,cm$^{-3}$ was searched for at 1.5 GHz using a 
maximum window size for matched-filtering of 130 ms. In the case of the coherently dedispersed 330~MHz data, a finer search was performed in the DM range  400--450 \,pc\,cm$^{-3}$, with a maximum window size of 260 ms. 
The pipeline finally performs sifting of the dedispersed data and produces plots for the surviving candidates.
Each candidate found by {\sc Heimdall} at DMs within the range
400--450\,pc\,cm$^{-3}$ was visually examined. 
In order to validate genuine candidates, we ran an ad-hoc program with more sensitive RFI excision procedures optimized for SRT data. This program searches for the most corrupted frequency channels in the DM zero data using the spectral kurtosis algorithm, with $5\sigma$ thresholding, as provided by the software package {\sc your} \footnote{\url{https://github.com/thepetabyteproject/your/}} \citep{Aggarwal2020}. It then applies baseline subtraction and normalizes the data for the average bandpass. A check for possible corrupted temporal bins due to the presence of impulsive RFI is then performed with inter-quantile range mitigation, similarly to \cite{10.1117/12.2559937}. Finally, it dedisperses the data to the derived DM and smooths them via a 2D Gaussian filter.
A single candidate not resembling RFI was found in the data at 1.5 GHz at   barycentric time
10:44:13.815 TDB (reported at infinite frequency) at a DM compatible with that of \frb:  DM = 421$\pm$ 4\,pc\,cm$^{-3}$. The burst had
a signal-to-noise ratio $S / N = 27.3$ and a width $W = 10$\,ms, resulting in a fluence $F=
13\pm3$\,Jy\,ms.
Since the data are uncalibrated, the estimated fluence of the burst was calculated using the
modified radiometer equation (see e.g. \citealt{lorimer04}) adopting an antenna gain of 0.55\,K/Jy,
a system temperature of 30\,K.
No burst was observed in the data at 330 MHz  either simultaneously with the burst at 1.5~GHz (with the time series both shifted  to infinite frequency in order to remove the delay introduced by DM at the two different frequencies), or at any other
time during the observing session, down to a limiting fluence of 7 Jy ms.

FAST carried out a 55-day observational campaign starting from April 1st 2021, soon after the CHIME alert \citep{ChimeAtel21}. More than 1600 FRB bursts were detected in 1.0-1.5 GHz frequency band. The details of FAST observations will be reported in a separate paper (H. Xu et al. 2021, in preparation).
During the \textit{Chandra} observation FAST  observed for 2 hours (8:00-10:00 UT) 
 detecting 48 bursts with fluences in the range 0.017-5.5 Jy ms  (median=0.22 Jy ms; average= 0.65 Jy ms) and  duration from 4 ms to 49 ms  (median=14 ms; average=19 ms).

We have searched for  coincidences  with  the 3 X-ray photons detected by \textit{Chandra}. 
The X-ray photon arrival times were barycenter corrected using the \texttt{barycorr} task within \texttt{HEASoft v.~6.27.2}.
The closest FRB-X-ray photon time difference is 314 s. Considering the number of burst and the count rate the probability of a random association is close to 100\%, thus we can exclude a significant coincidence with a FRB.
We derive a 5$\sigma$ upper limit on the X-ray fluence of the FRB as follows  \citep{Scholz17}.
For an X-ray burst arriving at the time of a FRB we obtain an upper limit of 14 counts (at 5 sigma \citep{Kraft1991} that corresponds to an upper limit of $\rm{F(2-10 keV)=2.5\times 10^{-10} erg\ cm^{-2}}$ on the fluence, and of E=$6\times 10^{45}$ erg on the X-ray burst energy, assuming the X-ray spectrum described in Appendix \ref{sec:Chandraimage}.  As the \textit{Chandra} background is negligible, this limit is independent of the duration of the FRB up to the difference in arrival time of the closest \textit{Chandra} photon  \citep{Scholz17}.
 In order to derive the upper limit for an X-ray burst occurring at a later time than 314 s after an FRB, we consider the number of trials and derive the single-trial confidence level that is used  to compute the upper limit following the prescription of \cite{Kraft1991}.  The number of  trials  is equal to the duration of the observation divided by the flare duration. Assuming a typical duration of 19 msec, comparable to the radio burst,  and considering the negligible background,  we derive an upper limit of 29 counts, corresponding to a fluence $\lesssim 5\times 10\rm{^{-10} erg\ cm^{-2}}$ and energy E$\lesssim 1.6\times 10^{46}$ erg).
 Under the assumption that X-ray bursts are emitted for each of the 49 radio bursts, we can stack individual limits to derive an upper limit of 0.3 cts. The corresponding upper limits on the fluence and energy of the putative X-ray bursts are thus  $5\times10^{-12} \rm{erg\,cm^{-2}}$ and $E_X$\,$\lesssim 1.1\times 10^{44}$ erg.
This corresponds to an upper limit on the ratio of the flare energy in  X-rays to radio $\rm{E_X/E_R\lesssim 5\times 10 ^{5}}$, and to a radio-to-X-ray spectral index  $\alpha_{RX}\gtrsim 0.4$ (F($\nu) \propto \nu^{-\alpha}$).
\begin{table}
\caption{Limits on X-ray bursts energy associated to FRBs}
\label{tab:FRB}
\begin{tabular}{lcccc}
 FRB  & E$_{XRB}$ & E$_{XRB,N}$ & N$_{FRB}$& Ref.  \\ \hline
	  	&	[erg]&	[erg] & & \\ \hline
20121102A	&$\lesssim	4\times 10^{46}$&$\lesssim	4\times 10^{45}$ & 10 & 1\\
20180916B &	$\lesssim 1.6\times 10^{45}$& $\lesssim	1.6 \times 10^{45}$ & 1& 2\\
20201124A&	$\lesssim	6\times 10^{45}$& $\lesssim	1.1 \times 10^{44}$ & 49 & 3\\
 \hline
\end{tabular}
\tablebib{(1)~\citet{Scholz17};
(2) \citet{Scholz20}; (3) this work}
\end{table}
In Tab.\ref{tab:FRB} we compare the  limits derived on the X-ray burst associated to FRBs with  two repeating FRBs (\frba; \citealt{Scholz17}, and, \frbb; \citealt{Scholz20}).
Overall, the most constraining observations for X-ray bursts in coincidence with FRB is E$_X\lesssim 1.6\times 10^{45}$ erg for a single burst in \frbb\ and E$_X\lesssim 1.1\times 10^{44}$ erg in \frb, assuming that all FRB have an associated XRB.

\end{appendix}









\end{document}